\documentclass[%
 reprint,
 amsmath,amssymb,
 aps,twocolumn,
prl,
]{revtex4-2}

\usepackage{graphicx}
\usepackage{dcolumn}
\usepackage{bm}
\usepackage[hypertexnames=false]{hyperref}
\usepackage{siunitx}
\usepackage[capitalise]{cleveref}
\usepackage{lipsum}

\usepackage{braket}
\newcommand{\tr}{\mathrm{tr}}

\usepackage{enumerate}
\usepackage{dsfont}

\usepackage{xcolor}

\usepackage[normalem]{ulem}

\begin{document}

\title{Surface Code Stabilizer Measurements for Rydberg Atoms}

\author{Sven Jandura}
\affiliation{University of Strasbourg and CNRS, CESQ and ISIS (UMR 7006), aQCess, 67000 Strasbourg, France}

\author{Laura Pecorari}
\affiliation{University of Strasbourg and CNRS, CESQ and ISIS (UMR 7006), aQCess, 67000 Strasbourg, France}

\author{Guido Pupillo}
\affiliation{University of Strasbourg and CNRS, CESQ and ISIS (UMR 7006), aQCess, 67000 Strasbourg, France}

\date{\today}

\begin{abstract}
We consider stabilizer measurements for surface codes with neutral atoms and identify gate protocols that minimize logical error rates in the presence of a fundamental error source -- spontaneous emission from Rydberg states. We demonstrate that logical error rates are minimized by protocols that prevent the propagation of Rydberg leakage errors and not by protocols that minimize the physical two-qubit error rate. We provide laser-pulse-level gate protocols to counter these errors.  These protocols significantly reduce the logical error rate for implementations of surface codes involving one or two species of atoms. Our work demonstrates the importance of optimizing quantum gates for logical errors in addition to gate fidelities and opens the way to the efficient realization of surface codes with neutral atoms. 
\end{abstract}

\maketitle

Quantum error correction (QEC) exploits redundancy to  encode a logical qubit with an arbitrarily low error rate into several noisy physical qubits, which is possible as long as the physical error rate is below a given threshold \cite{shor_scheme_1995, gottesman_stabilizer_1997, knill_theory_1997, lo_fault-tolerant_1998, steane_efficient_1999}. In surface codes \cite{bravyi_quantum_1998, dennis_topological_2002, fowler_surface_2012}, near faultless logical qubits are prepared and maintained by repeated measurement of so-called stabilizer operators on the physical qubits. In this way, up to $\lfloor d/2 \rfloor$ physical single qubit errors in $d$ rounds of stabilizer measurements can always be corrected, where $d$ is the so-called \emph{distance} of the code. Recently, several quantum computing platforms have surpassed the threshold for QEC by minimizing noise at the level of single- and two-qubit gates \cite{egan_fault-tolerant_2021, ryan-anderson_realization_2021, abobeih_fault-tolerant_2022, postler_demonstration_2022, evered_high-fidelity_2023, bluvstein_logical_2024, finkelstein_universal_2024,senoo2025highfidelityentanglementcoherentmultiqubit}. The errors on any quantum computing platform can be categorized as either errors in the computational subspace, or \emph{leakage} errors, where the qubit leaves the computational subspace. While errors in the computational subspace can be corrected with hardware-agnostic error correction codes, mitigating the effect of leakage errors requires hardware specific protocols \cite{werninghaus_leakage_2021, cong_hardware-efficient_2022, chow_circuit-based_2024}.
In neutral atom quantum computers \cite{saffman_quantum_2010, browaeys_many-body_2020, henriet_quantum_2020, morgado_quantum_2021, evered_high-fidelity_2023, bluvstein_logical_2024,graham_multi-qubit_2022, anand_dual-species_2024,scholl_erasure_2023,ma_high-fidelity_2023,cao_multi-qubit_2024, finkelstein_universal_2024}, 
two-qubit 
operations implemented by laser-exciting electrons to strongly interacting Rydberg states using time-optimal (TO) control methods \cite{jandura_time-optimal_2022,pagano_error_2022}
have recently surpassed gate fidelities $\mathcal{F}\simeq
99.5\%$ \cite{evered_high-fidelity_2023, finkelstein_universal_2024,senoo2025highfidelityentanglementcoherentmultiqubit}. These fidelities are above the threshold against errors in the computational subspace for QEC with surface- or color-codes, and  logical qubits and small logical circuits have been experimentally realized \cite{bluvstein_logical_2024}. However, a particularly detrimental error in this platform is Rydberg leakage, where an atom is erroneously left in a Rydberg state at the end of a gate. A Rydberg leakage error on a single qubit can prevent the execution of two-qubit gates on all of its neighbors, via the Rydberg blockade mechanism \cite{cong_hardware-efficient_2022, chow_circuit-based_2024}. It is thus a key open question how to perform stabilizer measurements that minimize logical error rates under realistic noise models including leakage errors.

\begin{figure*}
    \centering
    \includegraphics[width=\textwidth]{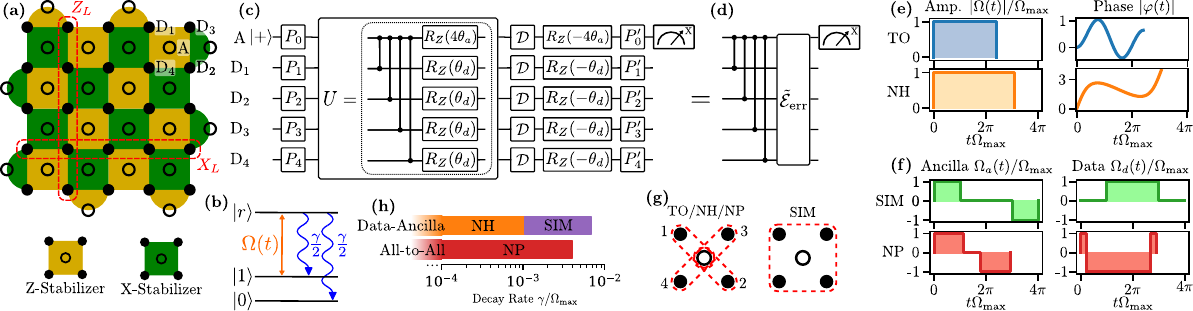}
    \caption{(a) A distance $d=5$ surface code with $d^2$ data qubit (solid circles) and $d^2-1$ ancilla qubits (open circles). $Z$-stabilizers are marked by yellow, $X$-stabilizers by green plaquettes. The logical operators $X_L$($Z_L$) consist of $X$($Z$) operators along a row(column) of data qubits. 
    (b) Each atom is modeled as a three level system with states $\ket{0},\ket{1},\ket{r}$, a time-dependent drive $\Omega(t)$ with phase $\varphi(t)$ and amplitude $|\Omega(t)| \leq \Omega_{\mathrm{max}}$ coupling $\ket{1}$ and $\ket{r}$, and a decay from $\ket{r}$ to $\ket{0}$ and $\ket{1}$ with decay rate $\gamma/2$ each.
    (c) Circuit to measure a $Z$-stabilizer (see main text).
    (d) The realistic stabilizer measurement is equivalent to four CZ gates followed by an error channel $\tilde{\mathcal{E}}_{\mathrm{err}}$. 
    (e) Amplitude and phase of $\Omega$ for the symmetric time-optimal (TO), no-hopping (NH) protocols.
    (f) Laser pulse $\Omega_a(t)$ and $\Omega_d(t)$ applied on the ancilla- and data atom respectively for the simultaneous (SIM) and no-phase (NP) protocols.  
    (g) For the TO/NH/NP protocols four CZ gates are applied subsequently in the indicated order. For the SIM protocol only one global gate is applied on all 5 atoms simultaneously.
    (h) Summary of which protocol has the lowest logical error rate as a function of $\gamma$ in the $d\rightarrow\infty$ limit.}
    \label{fig:intro}
\end{figure*}

In this work, we study how Rydberg decay, the dominant error sources in many experimental implementations \cite{evered_high-fidelity_2023, ma_high-fidelity_2023, graham_rydberg-mediated_2019} and a large source of Rydberg leakage errors, impacts the error rate of a logical qubit encoded in a surface code. We discuss two different architectures of neutral atom quantum computers, involving one or two species of stationary atoms as data and ancilla qubits, respectively. For both architectures we find that  for small enough Rydberg decay rates $\gamma$ the logical error rate is dominated by Rydberg leakage errors. The latter can cause errors in neighboring qubits during a stabilizer measurement, so that already $\lceil d/4 \rceil$ decay events are sufficient to cause a logical error. We propose and characterize pulse-level gate protocols that negate these errors by inhibiting the propagation of Rydberg excitations, so that again $\lfloor d/2 \rfloor$ errors can be corrected and the logical error rate is significantly reduced. Interestingly, these protocols differ significantly from those that minimize the average time that the atoms spend in the Rydberg state and thus the infidelity $1-\mathcal{F}$ of single two-qubit gates and stabilizer measurement. 
Our work demonstrates the need to optimize fidelities at the logical — not only the physical — level, and paves the way for efficient implementation of surface‑code quantum error correction on neutral‑atom platforms.

We consider a rotated surface code \cite{horsman_surface_2012} of distance $d$ made of $2d^2-1$ stationary atoms arranged in a square lattice [Fig.~\ref{fig:intro}(a)]. Stabilizers are given by products of Pauli $X$ or $Z$ operators on four data atoms on the corner of a plaquette, and read out using one ancilla qubit in the center of the plaquette. We assume that all stabilizers are measured  independently, and stabilizer measurements are only performed simultaneously if they involve disjoint sets of atoms. Let us  consider a single $Z$-stabilizer measurement ($X$-stabilizer measurements are performed in the same way, with Hadamard gates inserted on all data atoms before/after the measurement): Each of the five atoms on the plaquette is modeled as a three level system, with computational basis states $\ket{0}$, $\ket{1}$ and Rydberg state $\ket{r}$ [Fig.~\ref{fig:intro}(b)]. The dynamics is governed by the Hamiltonian ($\hbar=1$)
\begin{equation}
    H = \sum_{ij=0}^{4} B_{ij}\ket{r_ir_j}\bra{r_ir_j} + \sum_{i=0}^{4} \frac{\Omega_i(t)}{2}\ket{r_i}\bra{1_i} + \mathrm{h.c.}
    \label{eq:hamiltonian}
\end{equation}
where $B_{ij}$ is the interaction strength between atoms $i$ and $j$ prepared in the Rydberg state $\ket{r}$, $\Omega_i(t)$ is the (complex) time-dependent Rabi frequency of a laser incident on atom $i$ coupling $\ket{1}$ and $\ket{r}$. We consider either $B_{ij}=0$ or $B_{ij}=\infty$, corresponding, respectively, to the absence of interactions or a perfect {\it Rydberg blockade} between atoms $i$ and $j$, where simultaneous occupation of  Rydberg states for atoms $i$ and $j$ is impossible. Nevertheless, we show in the supplemental material \cite{sup_mat} that our results remain valid for gates implemented in the experimentally relevant strong but finite blockade regime.

In this work we adopt the simplest possible model of Rydberg decay, and assume that $\ket{r}$ decays only to the states $\ket{0}$ and $\ket{1}$ with a branching ratio of 1:1, described by the Lindblad operators $L_i^{(q)} = \sqrt{\gamma/2}\ket{q_i}\bra{r_i}$ for $i \in \{0,...,4\}$ and $q\in \{0,1\}$, where $\gamma$ is the decay rate of the Rydberg state. We show in the supplemental material \cite{sup_mat} that results do not change qualitatively for other branching ratios. In most experiments, there are additional decay channels to long-lived states outside of the computational subspace, which require repumping to the qubit states after the stabilizer measurement \cite{cong_hardware-efficient_2022}. 
Since for the purpose of a Rydberg blockade gate any low-lying non-computational state inhibits subsequent CZ operations and thus acts operationally like $\ket{0}$, these decay channels can be incorporated in our model by increasing the decay probability to $\ket{0}$. Moreover, through stimulated absorption or emission of black body radiation (BBR) it is also possible to cause the excitation of a \emph{different} Rydberg state. While not directly captured in our model, we show in the supplemental material \cite{sup_mat}  that in the presence of BBR decay the protocols developed in this work perform similarly to, and sometimes better than, standard protocols.

\begin{figure*}
    \centering
\includegraphics[width=\textwidth]{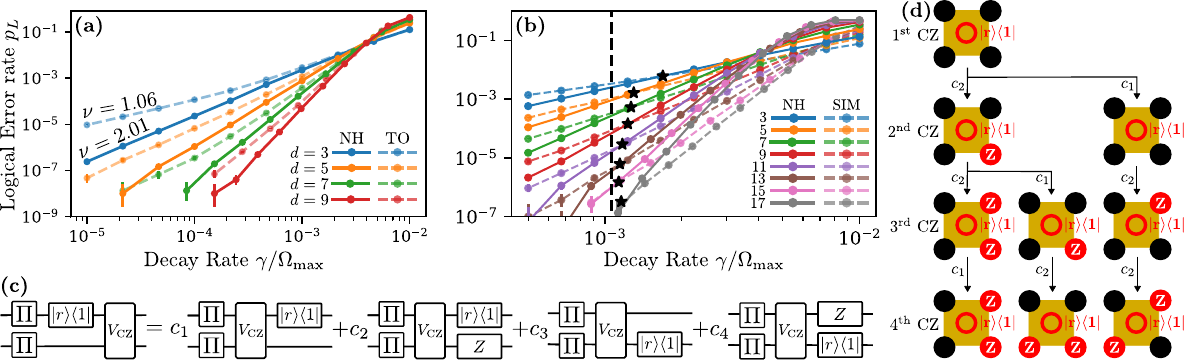}
    \caption{Data-ancilla blockade. 
    (a) Logical error rate $p_L$ as a function of the decay rate $\gamma$ (in units of $\Omega_{\mathrm{max}}$) for the NH (solid lines) and TO (dashed lines) protocols.
    (b) Logical error rate for the NH (solid lines) and SIM (dashed lines) protocol. Black stars show the crossover points between these two protocols. The black dashed line shows the asymptotic crossing point $\gamma_\times = 1.1\times 10^{-3}\Omega_{\mathrm{max}}$ in the limit $d\rightarrow\infty$ (see supplemental material \cite{sup_mat}). 
    (c) Four different ways how a $\ket{r}\bra{1}$ error can propagate through a $V_{\mathrm{CZ}}$ gate. Coefficients $c_1,..,c_4$ are calculated in the main text.  The projectors $\Pi$ restrict the state before the $\ket{r}\bra{1}$ error to the computational subspace.
    (d) Three possible ways how a $\ket{r}\bra{1}$ error on the ancilla qubit after the first gate (first row) can spread. The $n$-th row shows error after $n$ $V_{CZ}$ gates.}    \label{figv2:data_ancilla}
\end{figure*}

To measure a stabilizer, we follow the circuit shown in Fig.~\ref{fig:intro}(c): First, the ancilla atom is prepared in $\ket{+}=\left(\ket{0}+\ket{1}\right)/\sqrt{2}$. Then, laser pulses $\Omega_0$ and $\Omega_{1}(t)...,\Omega_{4}(t)$ are applied on the ancilla and the four data atoms, respectively, where we assume that for all pulses $|\Omega_j(t)| \leq \Omega_{\mathrm{max}}$, with $\Omega_{\mathrm{max}}$ the maximal achievable Rabi frequency. The pulses are chosen such that in the decay free case ($\gamma=0$) a unitary $U$ is implemented which, up to single qubit gates $R_Z(\theta_a) = \exp(i\theta_a \ket{1}\bra{1})$ and $R_Z(\theta_d)$ on ancilla and data atoms, respectively, corresponds to four controlled-Z (CZ) gates, one between each data atom and the ancilla atom (however, see discussion for the SIM protocol below). For simplicity we assume that all population remaining in the Rydberg state at the end of the pulses is then removed by applying the channel $\mathcal{D}(\rho) = \Pi \rho \Pi + \bra{r}\rho\ket{r}\Pi/2$ on each atom, with $\Pi = \ket{0}\bra{0}+\ket{1}\bra{1}$. This could be realized either by simply waiting long enough for the Rydberg state to decay with high probability, or by coupling the Rydberg state to a short lived intermediate state which decays to $\ket{0}$ and $\ket{1}$ \cite{cong_hardware-efficient_2022}. Thus, our model  captures the effect of Rydberg leakage errors during one stabilizer measurement, but not between different stabilizer measurements. The effect of leakage errors propagating across multiple stabilizer rounds can be readily included in our discussion and corrected using, e.g., leakage reduction units \cite{Perrin_2025,baranes2025,bluvstein2025architecturalmechanismsuniversalfaulttolerant} as we discuss in the supplemental material \cite{sup_mat}. Finally, single-qubit gates are applied on each atom to compensate for the single-qubit rotation induced by $U$, and the ancilla qubit is measured in the $X$ basis. The noisy stabilizer measurement can now be seen as four ideal CZ gates between the ancilla and the data qubits, followed by a 5-qubit error channel $\tilde{\mathcal{E}}_\mathrm{err}$ and the measurement of the ancilla qubit [Fig.~\ref{fig:intro}(d)]. We note that this equivalence is exact for the twirled noise channel. We indeed simulate $\tilde{\mathcal{E}}_\mathrm{err}$ using a Clifford simulator by applying randomized compiling \cite{wallman_noise_2016} and assuming that random Pauli gates $P$ and the Clifford conjugate gates $P'=UPU^\dagger$ are inserted before and after the unitary $U$, respectively. This ensures that $\tilde{\mathcal{E}}_\mathrm{err}$ is a Pauli channel, i.e. of the form
\begin{equation}
\tilde{\mathcal{E}}_{\mathrm{err}}(\rho) = \sum_Q \lambda_Q Q\rho Q \ 
\label{eq:error_channel}
\end{equation}
where $Q$ is summed over all 5-qubit Pauli strings and $\lambda_Q$ is the probability of Pauli error $Q$. We note that omitting $P$ and $P'$ from the measurement circuit while still assuming that $\tilde{\mathcal{E}}_{\mathrm{err}}$ is of the form of Eq.~\eqref{eq:error_channel} corresponds to the Pauli twirling approximation \cite{geller_efficient_2013}.
In the following, we compute  error probabilities $\lambda_Q$ by solving the Lindblad master equation given by $H$ and $L_i^{(q)}$. We then use the stabilizer circuit simulator STIM \cite{gidney_stim_2021} together with a minimum weight perfect matching decoder \cite{higgott_pymatching_2021} to calculate the logical error rate $p_L$ of $d$ rounds of stabilizer measurements using the computed $\lambda_Q$.

In order to exemplify the role of Rydberg leakage errors in the stabilizer measurements, 
we first discuss a so-called \emph{data-ancilla} blockade model, where we assume that for atoms on the same plaquette there is only a Rydberg blockade between the ancilla and all data atoms, but not between two data atoms. This model can be approximately realized in dual species arrays \cite{singh_dual-element_2022, anand_dual-species_2024, ireland_interspecies_2024}.  
We then discuss the \emph{all-to-all} blockade model, where we assume a Rydberg blockade between all atoms on a plaquette. This model can be approximately realized in single-species arrays \cite{graham_rydberg-mediated_2019, graham_multi-qubit_2022}.

In both blockade models, the stabilizer measurement can be realized by applying four subsequent CZ gates, each with the same pulse $\Omega_a(t)$ on the ancilla atom and $\Omega_d(t)$ on the data atom. The gate infidelity under Rydberg decay is minimized if each CZ gate is implemented using a time-optimal (TO) gate \cite{jandura_time-optimal_2022, pagano_error_2022}, in which the same laser pulse $\Omega(t)=\Omega_a(t)=\Omega_d(t)$ with constant amplitude $\Omega_{\mathrm{max}}$ and time dependent phase $\varphi(t)$ is applied \emph{symmetrically} on both atoms [Fig.~\ref{fig:intro}(e)]. 
This implements a $V_{CZ}:=\mathrm{CZ}\cdot[R_Z(\theta_a)\otimes R_Z(\theta_d)]$ gate corresponding to an exact CZ gate followed by single-qubit rotations  [where $R_Z(\theta_a)$ acts on the ancilla and $R_Z(\theta_d)$ acts on the data atom, respectively], with $\theta_a=\theta_d = 2.17$ \cite{jandura_time-optimal_2022}. 

Figure~\ref{figv2:data_ancilla}(a) shows the logical error rate $p_L$ for the TO protocol (dashed lines) for distances between $d=3$ and $d=9$ in the data-ancilla blockade model.  
In particular, for $d=3$ we find that for small enough decay rates $\gamma$ the logical error rate scales as $p_L \sim \gamma^\nu$ with  $\nu \approx 1$, meaning that already a single decay event can cause a logical error. This contradicts the expectation that a $d=3$ surface code should be able to correct a single decay event. The same trend persists at larger distances. The $\nu$ values extracted from the fits of displayed higher-distance decoding curves are: $\nu_{TO}^{d=5}=2.12$ and $\nu_{NH}^{d=5}=2.87$; $\nu_{TO}^{d=7}=2.90$ and $\nu_{NH}^{d=7}=3.83$; $\nu_{TO}^{d=9}=4.10$ and $\nu_{NH}^{d=9}=4.96$.

To understand why the TO protocol performs worse than expected, it is sufficient to consider a single decay event during the first of the four $V_{\mathrm{CZ}}$ gates: Due to the fraction of the laser pulse that is applied after the decay event it is possible that the ancilla qubit is found in $\ket{r}$ at the end of the gate. Instead of a Pauli $X$, $Y$ or $Z$ error this corresponds to a $\ket{r}\bra{1}$ error which occurs right after the first $V_{\mathrm{CZ}}$ gate on the ancilla qubit. 
This latter type of Rydberg error can be detrimental, as $V_{CZ}$ only acts as a CZ gate on the computational subspace, but its action on states outside of the computational subspace depends on the gate protocol. In particular, we find that a $\ket{r}\bra{1}\otimes I$ error followed by a $V_{\mathrm{CZ}}$ gate is equivalent to a $V_{\mathrm{CZ}}$ gate followed by a linear combination of $\ket{r}\bra{1}\otimes I$, $\ket{r}\bra{1}\otimes Z$, $I\otimes \ket{r}\bra{1}$ and $Z\otimes \ket{r}\bra{1}$ errors, with coefficients $c_1,c_2,c_3$ and $c_4$, respectively [Fig~\ref{figv2:data_ancilla}(c)].
In the supplemental material \cite{sup_mat}, the coefficients $c_1$ to $c_4$ for a symmetric protocol with $\theta_a=\theta_d=:\theta$ in which the same pulse $\Omega(t)$ is applied to both atoms are calculated as 
\begin{eqnarray}
    c_{1/2} &=& \left(e^{-2i\theta} \mp ie^{-3i\theta}\sin\theta\right)/2 \label{eq:c12} \\
    c_{3/4} &=& \pm\left(e^{-3i\theta}\cos\theta\right)/2 \label{eq:c34}
\end{eqnarray}
where the $-$ sign is used for $c_1$ and $c_4$ and the $+$ sign is used for $c_2$ and $c_3$. In particular, for the TO protocol we find that all $c_i$ are nonzero and thus all four propagation channels occur with a nonzero amplitude. Hence, a single $\ket{r}\bra{1}$ error after the first $V_{\mathrm{CZ}}$ gate can propagate in $3^4$ different ways. Three of those error propagation paths are shown in Fig.~\ref{figv2:data_ancilla}(d), where each of the four rows shows possible errors after each of the four $V_{\mathrm{CZ}}$ gates.  We see that a single $\ket{1}\bra{r}$ error on the ancilla qubit can lead to pairs of $Z$ errors on the data qubits that are aligned horizontally, vertically, or diagonally. The pair of vertically aligned $Z$ errors is particularly detrimental, since the logical $Z$ operator is oriented vertically [Fig.~\ref{fig:intro}(a)] and thus only $\lceil d/4 \rceil$ of those errors suffice to make a logical $Z$ error. This is in contrast to independent single qubit errors, where $\lceil d/2 \rceil$ errors are necessary for a logical error.  Note that, since also horizontally or diagonally aligned errors can occur, this behavior cannot be fixed by changing the order in which the $V_{\mathrm{CZ}}$ gates are applied to the data qubits. 

Equations~\eqref{eq:c12} and~\eqref{eq:c34} show however a way in which the spreading of the $\ket{1}\bra{r}$ error can be prevented: For $\theta=\pi/2$ we obtain $c_1=-1, c_2=c_3=c_4=0$, so that the $\ket{1}\bra{r}$ error commutes with the $V_{\mathrm{CZ}}$ gate, up to a global phase. Using quantum optimal control methods \cite{khaneja_optimal_2005} and the methodology developed in \cite{jandura_time-optimal_2022} we determine the shortest possible pulse $\Omega_1(t)=\Omega_2(t)$ that implements a $V_{\mathrm{CZ}}$ gate with $\theta=\pi/2$ [see Fig.~\ref{fig:intro}(e)]. Similar to the TO pulse \cite{jandura_time-optimal_2022}, this pulse requires a continuous variation of the laser phase with time only, while the amplitude is kept fixed at $\Omega_{\rm max}$, which is expected to be experimentally advantageous \cite{evered_high-fidelity_2023}. When applied to the whole plaquette, this continuous pulse forms the basis of the \emph{No-Hopping (NH)} protocol for stabilizer measurements. Note that in the NH protocol, the time spent in the Rydberg state is increased by less than 10\% compared to the TO protocol. The solid lines in Fig.~\ref{figv2:data_ancilla}(a) show the logical error rate for the NH protocol: For large decay rates $\gamma \gtrsim 10^{-3}\Omega_{\mathrm{max}}$ the NH protocol yields essentially the same logical error rates as the TO protocol since both protocols spend a similar time in the Rydberg state. However, for small decay rates $\gamma \lesssim 10^{-3}\Omega_{\mathrm{max}}$, the NH protocol significantly reduces the logical error rate for all distances $d$, demonstrating that in this regime the propagation of Rydberg leakage errors becomes the dominant error source. For $d=3$, $\nu$ takes the expected value $\nu = 2.01 \approx 2$, showing that two decay events are necessary to produce one logical error in the NH protocol. We expect that in  the currently experimentally accessible region $10^{-4} \Omega_{\mathrm{max}} \lesssim \gamma\lesssim 10^{-3} \Omega_{\mathrm{max}}$\cite{evered_high-fidelity_2023,anand_dual-species_2024,scholl_erasure_2023,ma_high-fidelity_2023,cao_multi-qubit_2024} the logical error rate can then be improved by more than one order of magnitude. Since the exponent $\nu$ scales linearly with the distance $d$ (see supplemental material \cite{sup_mat}) and the logical error rate scales as $p_L\sim\gamma^{\nu}$, at fixed $0<\gamma<1$ even larger improvements are expected for larger distances. Finally, we note that, although the NH derivation assumes an ideal infinite blockade (so that $\theta=\pi/2$ ensures $c_1=-1,c_2=c_3=c_4=0$), we show in the supplemental material \cite{sup_mat} that the protocol’s favorable logical-error scaling persists for experimentally relevant, strong but finite blockade strengths such as $B=20\Omega_\text{max}$.

The data-ancilla blockade also allows for a stabilizer measurement protocol  (called SIM protocol from now on) in which the four CZ gates are not applied subsequently, but \emph{simultaneously}, by applying a $\pi$-pulse on the ancilla qubit, followed by a $2\pi$-pulse applied on all data qubits, and a final $\pi$-pulse on the ancilla qubit [Fig. ~\ref{fig:intro}(f)]. This is similar to the original proposal by Ref. \cite{jaksch_fast_2000} for Rydberg blockade gates, though here the $2\pi$-pulse is applied to all qubits, which reduces the time spent in the Rydberg state by 
$\sim 40\%$ compared to the NH protocol. Fig.~\ref{figv2:data_ancilla}(b) shows that for a decay rate $\gamma > \gamma_\times = 1.1\times 10^{-3}\Omega_{\mathrm{max}}$ the SIM protocol also reduces the logical error rate compared to the NH protocol. However, in the SIM protocol a single decay event of the ancilla qubit during the $2\pi$ pulse on the data qubits can lead to \emph{any} 5-qubit error $Q$, so that again only $\lceil d/4 \rceil$ decay events are necessary for a logical error (see supplemental material~\cite{sup_mat}). We thus find that for $\gamma < \gamma_\times$ the NH protocol achieves the lower logical error rate [see Fig.~\ref{fig:intro}(h)].

\begin{figure}[t]
    \centering
    \includegraphics{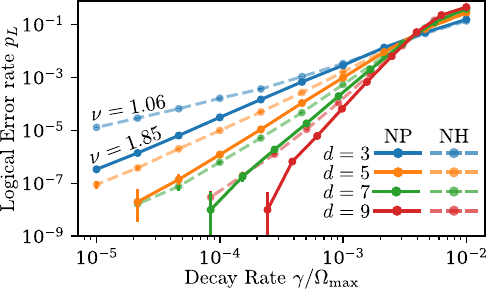}
    \caption{All-to-all blockade. 
    Logical error rate $p_L$ as a function of the decay rate $\gamma$ (in units of $\Omega_{\mathrm{max}}$) for the NP (solid lines) and NH (dashed lines) protocols.}
    \label{figv2:all_to_all}
\end{figure}

Finally we briefly turn to the all-to-all blockade model, in which there is also a Rydberg blockade between data qubits on the same plaquette. In this case, we find that the NH protocol is not sufficient to restore the $\lceil d/2 \rceil$ scaling [see Fig.~\ref{figv2:all_to_all}]. Similar to the data-ancilla case, we show that this behavior is due to the propagation of Rydberg leakage errors on data qubits, and can only be prevented through a gate protocol with $\theta_d \in \{0,\pi\}$ (see supplementary material \cite{sup_mat}). This is only possible with an \emph{asymmetric} protocol ($\Omega_a \neq \Omega_d$), since we saw above that symmetric protocols have to satisfy $\theta_a=\theta_d=\pi/2$. Using quantum optimal control we find the shortest protocol with \emph{i)} $\theta_d\in \{0,\pi\}$ and \emph{ii)} $c_1=1, c_2=c_3=c_4=0$, called No-Phase (NP) protocol [see Fig.~\ref{fig:intro}(f)], which indeed restores the expected $\lceil d/2 \rceil$ scaling [see Fig.~\ref{figv2:all_to_all}]. Consistently, the slopes $\nu$ extracted from the fits of displayed decoding curves are: $\nu_{NP}^{d=3}=1.85$ and $\nu_{TO}^{d=3}=1.06$; $\nu_{NP}^{d=5}=2.95$ and $\nu_{TO}^{d=5}=2.14$; $\nu_{NP}^{d=7}=3.92$ and $\nu_{TO}^{d=7}=2.83$; $\nu_{NP}^{d=9}=5.05$ and $\nu_{TO}^{d=9}=3.48$.  We note that also the original $\pi$-$2\pi$-$\pi$ protocol \cite{jaksch_fast_2000} satisfies conditions \emph{i)} and \emph{ii)}, but has a 30\% longer Rydberg time and thus a larger logical error rate than the NP protocol (see \cite{sup_mat}).

In conclusion, we have established that, due to the propagation of Rydberg leakage errors, gate protocols which minimize the two-qubit gate infidelity do not necessarily achieve the lowest possible logical error rates. For the data-ancilla and all-to-all blockade model we identified the relevant error mechanisms and mitigated them by introducing two new protocols for performing stabilizer measurements with Rydberg atoms, which significantly reduce the logical error rate. Interestingly both protocols require breaking the symmetry between ancilla and data atoms, either by an asymmetric blockade model or by an asymmetric laser pulse. Similar results are also expected for other quantum error correction codes, such as LDPC codes, when implemented on stationary arrays of neutral atoms \cite{pecorari_high-rate_2024}.
We expect that our protocols also mitigate Rydberg leakage errors arising due to other errors sources, such as Rydberg dephasing. The precise effect of this and other error sources will be subject of future work. 
Thus, in a broader context and in line with the goals of fault-tolerant quantum computing, our work demonstrates the necessity to optimize stabilizer measurements beyond two-qubit gate fidelities to achieve the lowest possible logical error rate. This requirement becomes increasingly urgent as platforms approach $99.9\%$ two-qubit gate fidelities and efforts intensify to reduce QEC cycle times from milliseconds to microseconds, a regime in which the error mechanisms analyzed in this work are expected to become dominant.

\begin{acknowledgements}
\it
We are grateful to Jeff Thompson, Shruti Puri, Shannon Whitlock and Asier
Piñeiro Orioli for stimulating discussions. This research received funding from the European Union’s Horizon 2020 program under the Marie Sklodowska-Curie project 955479 (MOQS), the Horizon Europe program HORIZON-CL4-2021- DIGITALEMERGING-01-30 via the project 101070144 (EuRyQa) and from the French National Research Agency under the Investments of the Future Program project ANR-21-ESRE-0032 (aQCess), ANR-17-EURE-0024 (QMat), and ANR-22-CMAS-0001 France 2030 (QuanTEdu-France) and the Institut Universitaire de France (IUF).
\end{acknowledgements}

\bibliography{library}

\onecolumngrid
\clearpage
\begin{center}
\textbf{\large Supplementary Material for ``Surface Code Stabilizer Measurements for Rydberg Atoms"}

\vspace{0.25cm}

Sven Jandura, Laura Pecorari, and Guido Pupillo

\emph{University of Strasbourg and CNRS, CESQ and ISIS (UMR 7006), aQCess, 67000 Strasbourg, France}

{\small (Dated: \today)}
\end{center}
\maketitle
\setcounter{equation}{0}
\setcounter{figure}{0}
\setcounter{table}{0}
\makeatletter
\renewcommand{\thesection}{S\thesection.\arabic{section}}
\renewcommand{\theequation}{S\arabic{equation}}
\renewcommand{\thefigure}{S\arabic{figure}}
\twocolumngrid 

\section{S1. Numerical calculation of error probabilities}
To calculate the error probabilities $\lambda_Q$ in Eq. (2) [main text] we first numerically integrate the Lindblad master equation given by $H$ and the $L^{(q)}_i$ with the initial condition $\rho(0)=R$ for each 5 qubit Pauli string $R$, obtaining the result $\mathcal{F}(R) := \rho(T)$, where $T$ is the duration of the laser pulses. We then calculate the error channel $\mathcal{E}_{\mathrm{err}}$ , without the application of the Pauli gates used for randomized compiling, as
\begin{equation}
    \mathcal{E}_{\mathrm{err}}(R) = U^\dag \mathcal{D}^{\otimes 5}(\mathcal{F}(R))U.
    \label{eq:SI_E_err}
\end{equation}
$\mathcal{E}_{\mathrm{err}}$ can be expressed using its $\chi$ matrix as 
\begin{equation}
    \mathcal{E}_{\mathrm{err}}(\rho) = \sum_{Q,Q'} \chi_{Q,Q'} Q\rho Q'
    \label{eq:SI_chi_matrix}
\end{equation}
where $Q$ and $Q'$ are summed over all 5 qubit Pauli strings. The twirled channel is then given by \cite{geller_efficient_2013}
\begin{equation}
    \tilde{\mathcal{E}}_{\mathrm{err}}(\rho) := 4^{-5}\sum_{P}P\mathcal{E}_{\mathrm{err}}(P\rho P)P = \sum_Q \lambda_Q Q\rho Q
\end{equation}
where $\lambda_Q = \chi_{Q,Q}$. 

Now we use that 
\begin{eqnarray}
    \label{eq:SI_trRER}
    \tr(R\mathcal{E}_{\mathrm{err}}(R)) &=& \sum_{Q,Q'}\chi_{Q,Q'}\tr(RQRQ')\\ 
    &=& \nonumber \sum_Q \lambda_Q \tr((RQ)^2)\\
    &=& \nonumber 2^5 \sum_Q s(R,Q) \lambda_Q
\end{eqnarray}
where the second equality follows from $\tr(RQRQ') = \pm \tr(QQ') = \pm 2^5 \delta_{QQ'} $ and $s(R,Q) = 1$ if $R$ and $Q$ commute and $s(R,Q) = -1$ if $R$ and $Q$ anti-commute.

Inverting Eq.~\eqref{eq:SI_trRER} gives
\begin{equation}
    \lambda_Q = 4^{-5} \sum_R s(R,Q) \tr(R\mathcal{E}_{\mathrm{err}}(R))
\end{equation}
so $\lambda_Q$ can be calculated using Eq.~\eqref{eq:SI_E_err}.

\section{S2. Decomposition of error probabilities into decay events}
In the following we calculate $\lambda_Q$ to first order in $\gamma$. For this, denote by $U(t_1, t_2)$ the evolution operator under $H(t)$ (without Rydberg decay) from time $t=t_1$ to $t=t_2$. Note that $U = U(0,T)$. For an initial density matrix $\rho(0)$, the solution of the Lindblad equation is to first order in $\gamma$ given by
\begin{align}
    \rho(T) = U\rho(0)U^\dag + \sum_{q,i}&\int_0^T\mathrm{d}t  \Big[  F^{(q)}_{i}(t)\rho(0)F^{(q)}_{i}(t)^\dag \\ \nonumber
    &- \left\{F^{(q)}_{i}(t)^\dag F^{(q)}_{i}(t), \rho(0)\right\}/2\Big]
\end{align}
with $F^{(q)}_{i}(t) = U(t,T)L^{(q)}_{i}U(0,t)$. 

Thus, the error channel $\mathcal{E}_{\mathrm{err}}$ (see Eq.~\eqref{eq:SI_E_err}) is given by
\begin{align}
    \label{eq:Si_E_err_full}
    \mathcal{E}_{\mathrm{err}}(\rho) = \rho + \sum_{q,i,k}&\int_0^T\mathrm{d}t  \Big[  E^{(q)}_{i,k}(t)\rho E^{(q)}_{i,k}(t)^\dag \\ \nonumber
    &- \left\{E^{(q)}_{i,k}(t)^\dag E^{(q)}_{i,k}(t), \rho \right\}/2\Big]
\end{align}
with
$E^{(q)}_{i,k} = U^\dag D_k U(t,T)L^{(q)}_{i}U(0,t)$ where the $D_k$ are the Kraus operators of $\mathcal{D}^{\otimes 5}$.

Now every $E^{(q)}_{i,k}(t)$ can be expanded as 
\begin{equation}
    E^{(q)}_{i,k}(t) = 2^{-5}\sum_Q \tr\left(QE^{(q)}_{i,k}(t)\right)Q
\end{equation}
where $Q$ is summed over all 5-qubit Pauli strings. From comparing Eq.~\eqref{eq:Si_E_err_full} with Eq.~\eqref{eq:SI_chi_matrix} we can read off that for $Q\neq I^{\otimes 5}$
\begin{align}
    \lambda_Q = \chi_{Q,Q} = 4^{-5} \sum_{q,i,k}&\int_0^T\mathrm{d}t \left|\tr\left(QE^{(q)}_{i,k}(t)\right)\right|^2
    \label{eq:SI_lambda_Q}
\end{align}

The interpretation of Eq.~\eqref{eq:SI_lambda_Q} is that in order to find the error probabilities $\lambda_Q$, at least to first order in $\gamma$, it is sufficient to consider single decay events. Each decay event is associated with a Lindblad operator $L^{(q)}_i$, specifying that qubit $i$ decays from state $\ket{r}$ to state $\ket{q}$, with a time $t$ at which the decay takes place, and with a Kraus operator $D_k$ specifying how the channel $\mathcal{D}^{\otimes 5}$ acts. The index $k=(k_0,...,k_4)$ is summed over $\{0,1,2\}^{5}$, and the Kraus operators are given by $D_k = \bar{D}_{k_0}\otimes...\otimes \bar{D}_{k_4}$ with $\bar{D}_0 = \ket{0}\bra{r}/\sqrt{2}$, $\bar{D}_1 = \ket{1}\bra{r}/\sqrt{2}$ and $\bar{D}_2 = \Pi = \ket{0}\bra{0}+\ket{1}\bra{1}$.

In the main text and below, we show that for error strings $Q$ with two vertically aligned $Z$ errors on the data qubits we obtain $\lambda_Q = \Theta(\gamma)$ (i.e. $\lambda_Q$ is of order $\gamma$) by considering a single decay event and showing that $\tr\left(QE^{(q)}_{i,k}(t)\right) \neq 0$. Since all terms in Eq.~\eqref{eq:SI_lambda_Q} are nonnegative, this is indeed sufficient to show that $\lambda_Q = \Theta(\gamma)$.

\section{S3. Propagation of $\ket{r}\bra{1}$ errors under $V_{\mathrm{CZ}}$}
\subsection{S3.1. Symmetric protocols}
In the following we derive Eqs.(3) and (4) from the main text. For this we consider a symmetric protocol, i.e. a protocol in which the same pulse $\Omega(t)$ is applied to both atoms. The action of such a pulse can be understood by considering the two two-level systems \cite{levine_parallel_2019, jandura_time-optimal_2022}
\begin{equation}
    H_1 = \frac{\Omega(t)}{2}\ket{0r}\bra{01}+\mathrm{h.c.}
\end{equation}
and 
\begin{equation}
    H_2 = \frac{\sqrt{2}\Omega(t)}{2}\ket{W_+}{\langle11|}+\mathrm{h.c.}
\end{equation}
with $\ket{W_\pm} = (\ket{1r}\pm\ket{r1})/\sqrt{2}$. $V_{CZ}$ can be obtained by integrating the Schrödinger equation with $H_1$ and $H_2$.

The pulse $\Omega(t)$ is chosen such that $V_{CZ}$ acts as $V_{CZ}\ket{01} = e^{i\theta}\ket{01}$ and $V_{CZ}\ket{11} = -e^{2i\theta}\ket{11}$. Since $H_1$ and $H_2$ are traceless it follows that $V_{CZ}$ must have determinant 1 when restricted to the $\{\ket{01}, \ket{0r}\}$ or $\{\ket{11}, \ket{W_+}\}$ subspaces. Thus it follows that $V_{CZ}\ket{0r} = e^{-i\theta}\ket{0r}$ and $V_{CZ}\ket{W_+} = -e^{-2i\theta}\ket{W_+}$. Additionally, since the same pulse is applied to both atoms but $\ket{W_-}$ is antisymmetric under the exchange of the atoms, $\ket{W_-}$ is unaffected by the pulse, i.e $V_{CZ}\ket{W_-} = \ket{W_-}$. Given the evolution of $\ket{W_\pm}$ under $V_{CZ}$ we can now compute
\begin{align}
    V_{CZ}\ket{1r} &= (V_{CZ}\ket{W_+}+V_{CZ}\ket{W_-})/\sqrt{2} \\ \nonumber
    &=  e^{-i\theta}\left(i\sin(\theta)\ket{1r} - \cos(\theta)\ket{r1}\right)
\end{align}
and analogously
\begin{align}
     V_{CZ}\ket{r1} =  e^{-i\theta}\left(i\sin(\theta)\ket{r1} - \cos(\theta)\ket{1r}\right).
\end{align}

Now we want to find an error $E$ such that applying $V_{CZ}$ after a $\ket{r}\bra{1}$ error is equivalent to applying first $V_{CZ}$ and then $E$, at least under the condition that the before the $\ket{r}\bra{1}$ error we are in the computational subspace. Formally, we want to find $E$ such that
\begin{equation}
    V_{CZ}(\ket{r}\bra{1} \otimes I)\Pi^{\otimes 2} = EV_{CZ}\Pi^{\otimes 2}
    \label{eq:SI_VCz_propagation}
\end{equation}
Using that $[V_{CZ},\Pi^{\otimes 2}]=0$ and $(\ket{r}\bra{1}\otimes I)\Pi^{\otimes 2} = \ket{r0}\bra{10} + \ket{r1}\bra{11}$ we obtain
\begin{align}
\label{eq:SI_EPi_full}
    E\Pi^{\otimes 2} &= V_{CZ}\ket{r0}\bra{10}V_{CZ}^\dag + V_{CZ}\ket{r1}\bra{11}V_{CZ}^\dag \\ \nonumber
    &= e^{-2i\theta}\ket{r0}\bra{10} - e^{-3i\theta} [\sin(\theta)\ket{r1}\bra{11} \\ \nonumber 
    &- \cos(\theta)\ket{1r}\bra{11}] \\ \nonumber
    &= e^{-2i\theta}\ket{r}\bra{1} \otimes (I+Z)/2\\ \nonumber & -ie^{-3i\theta}\sin(\theta) \ket{r}\bra{1} \otimes (I-Z)/2 \\ \nonumber
    &+e^{-3i\theta}\cos(\theta) (I-Z)/2 \otimes \ket{r}\bra{1}
\end{align}

For $E = c_1 \ket{r}\bra{1} \otimes I + c_2 \ket{r}\bra{1} \otimes Z + c_3 I \otimes \ket{r}\bra{1} + c_4 Z \otimes \ket{r}\bra{1}$ Eq.~\eqref{eq:SI_VCz_propagation} is indeed satisfied with $c_1,c_2,c_3$ and $c_4$ given by Eq. (3) and (4).

\subsection{S3.2. The $\pi$-$2\pi$-$\pi$ protocol}
For completeness we show here that the $\pi$-$2\pi$-$\pi$ protocol \cite{jaksch_fast_2000} does not spread Rydberg excitation errors. On states outside of the computational subspace, the $\pi$-$2\pi$-$\pi$ protocol acts as follows (we assume the first qubit is the ancilla qubit, on which the two $\pi$ pulses are applied, and the second the data qubit, on which the $2\pi$ pulse is applied);

\begin{align}
    V_{CZ}\ket{0r} &= - \ket{0r} \\
    V_{CZ}\ket{r0} &= \ket{r0} \\
    V_{CZ}\ket{1r} &=  -\ket{1r} \\
    V_{CZ}\ket{r1} &= -\ket{r1}
\end{align}

Thus we obtain for a $\ket{r}\bra{1}$ error on the ancilla qubit
\begin{align}
    V_{CZ}(\ket{r}\bra{1}\otimes I)\Pi^{\otimes 2} &= \ket{r0}\bra{10} + \ket{r1}\bra{11}\\ \nonumber 
    &= (\ket{r}\bra{1}\otimes I) V_{CZ} \Pi^{\otimes 2}
\end{align}
corresponding to $c_1=1$ and $c_2=c_3=c_4 = 0$.

\begin{figure}
    \centering
    \includegraphics[width=\linewidth]{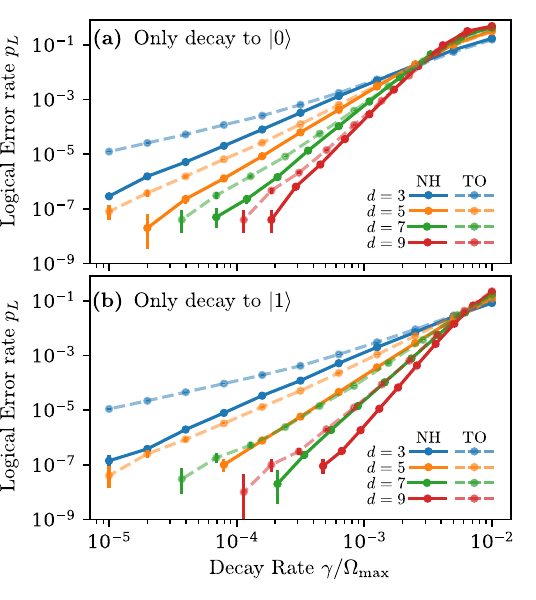}
    \caption{Data-ancilla blockade with different branching ratios. (a) Logical error rate $p_L$ as a function of the decay rate $\gamma$ (in units of $\Omega_{\mathrm{max}}$) for the NH (solid lines) and TO (dashed lines) protocols assuming only Rydberg decays to $|0\rangle$. (b) Logical error rate $p_L$ as a function of the decay rate $\gamma$ (in units of $\Omega_{\mathrm{max}}$) for the NH (solid lines) and TO (dashed lines) protocols assuming only Rydberg decays to $|1\rangle$.}
    \label{fig:SI_branchings}
\end{figure}

\section{S4. Results for other Branching Ratios}

In the following, we discuss the choice of a 1:1 branching ratio for Rydberg decays made in the main text and comment on other branching ratios. 
 
In this work we have adopted the simplest possible model of Rydberg decay and assumed that $\ket{r}$ decays only to the states $\ket{0}$ and $\ket{1}$ in equal fraction, corresponding to a branching ratio of 1:1. We observe that this assumption is not restrictive and successes to capture the qualitative behavior of arbitrary leakage errors within a single stabilizer round. That is because, although atoms can decay to other long-lived states (e.g., to other hyperfine ground states out of the qubit subspace), any low-lying state outside of the computational space essentially behaves like $\ket{0}$. The state $\ket{0}$ is indeed dark to the Rydberg laser, hence preventing the execution of a CZ gate as if the atoms was missing. Therefore, these decay channels can be accounted for in our model by simply modifying the branching ratio and increasing the probability of decay to $\ket{0}$. In this regard, we observe that the qualitative behaviors of the Time-Optimal (TO) and No-Hopping (NH) protocols do not change with other branching ratios when the four CZ gates are applied sequentially as modeled in the main text. As an example, we show in Fig.~\ref{fig:SI_branchings}(a),(b) the surface code simulations for TO and NH protocols assuming only decay to $|0\rangle$ and $|1\rangle$, respectively. These findings demonstrate that the NH protocol still outperforms the TO one by preserving a favorable scaling of the logical error rate, thus further reinforcing the results derived in the main text. We discuss more in detail below (in supplemental section XI) hardware-level and circuit-level strategies to prevent the propagation of leakage errors across multiple stabilizer rounds and comment on how these could be readily integrated with our noise model to deliver a comprehensive description of leakage errors in neutral atom quantum processors.

\section{S5. Finite blockade interaction strength}
In the following, we show that the claims of the main text derived under the assumption of infinite blockade strength hold also for strong but finite blockade with good approximation. We here focus on the NH protocol, although most of the considerations can be generalized to the other cases as well.

We consider again the propagation of $|r\rangle\langle 1|$ errors through $V_{CZ}$ for symmetric protocols, such as the NH protocol. If the blockade interaction strength between the two atoms is finite, the action of the pulse $\Omega(t)$ can now be understood by considering the following two two-level systems:
\begin{equation}
    H_1 = \frac{\Omega(t)}{2}\ket{0r}\bra{01}+\mathrm{h.c.}
\end{equation}
and
\begin{equation}
    H_2 = \frac{\sqrt{2}\Omega(t)}{2}\ket{W_+}\langle11|+\mathrm{h.c.}+B|rr\rangle\langle rr|
\end{equation}
with $\ket{W_\pm} = (\ket{1r}\pm\ket{r1})/\sqrt{2}$. As before, the pulse $\Omega(t)$ is chosen such that $V_{CZ}$ acts as $V_{CZ}\ket{01} = e^{i\theta}\ket{01}$ and $V_{CZ}\ket{11} = -e^{2i\theta}\ket{11}$. However, now only $H_1$ is traceless, hence it follows that $V_{CZ}$ must have determinant $1$ only when restricted to the $\{\ket{01}, \ket{0r}\}$ subspace. Instead, $\text{Tr}(H_2)=B$, hence $\text{det}(V_{CZ})=\exp[i\text{Tr}(H_2)t]=e^{iBt}$. Thus, it follows that $V_{CZ}\ket{0r} = e^{-i\theta}\ket{0r}$ and $V_{CZ}\ket{W_+} = -e^{-2i\theta+iBt}\ket{W_+}$. 
We observe that the term $e^{iBt}$ oscillates rapidly for large $B$, so its contribution averages out over typical gate durations and can be neglected (analogously to employing the
rotating wave approximation). Consequently, the NH protocol’s ability to suppress error correlations remains robust as long as the blockade interaction strength significantly exceeds the maximum Rabi frequency, $B \gg \Omega_\text{max}$. In contrast, for $B \sim \Omega_\text{max}$, the propagation of $|r\rangle\langle 1|$ through $V_{CZ}$ changes qualitatively, and the NH protocol can no longer ensure a favorable logical error scaling, necessitating the development of new, pulse-specific error suppression strategies.

\begin{figure}[t!]
    \centering
    \includegraphics[width=\linewidth]{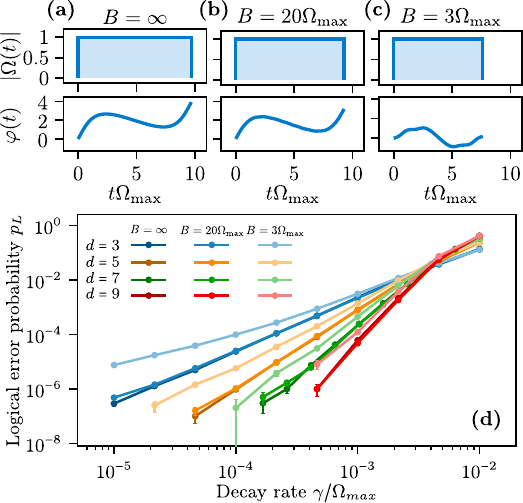}
    \caption{NH protocol at infinite and finite blockade strength. (a)--(c) optimal amplitude and phase profiles for the NH protocol ($\theta=\pi/2$) with $B=\infty,20\Omega_\text{max},3\Omega_\text{max}$, respectively. (d) Logical error rate $p_L$ as a function of the decay rate $\gamma$ (in units of $\Omega_\text{max}$) for the NH protocol at infinite and finite blockade strength. In the strong blockade limit, the favorable logical error rate scaling is maintained.}
    \label{fig:NH_finiteB}
\end{figure}

To validate our claims, we derive the pulses for the NH protocol at finite blockade strength with $B=20\Omega_\text{max}$ and $B=3\Omega_\text{max}$, and perform surface code numerical simulations in these two cases. Fig.~\ref{fig:NH_finiteB}(a)--(c) show the optimal pulses with fixed $\theta=\pi/2$ for $B=\infty$, $B=20\Omega_\text{max}$ and $B=3\Omega_\text{max}$. We note that for $B=20\Omega_\text{max}$ the pulse phase profile is qualitatively similar to that for $B=\infty$. Moreover, in contrast to the standard NH protocol with pulse duration $T\Omega_{\text{max}} = 9.7$, the optimized pulses at finite blockade are shorter, reflecting the increased population of Rydberg-excited states. That is, $T\Omega_{max}=9.3$ and $T\Omega_{max}=7.6$ for $B=20\Omega_\text{max}$ and $B=3\Omega_\text{max}$, respectively. Fig.~\ref{fig:NH_finiteB}(d) shows instead the results of surface code numerical simulations for the NH protocol with $B=\infty$, $B=20\Omega_\text{max}$, and $B=3\Omega_\text{max}$. We observe that the logical error rate of the NH protocol with $B=20\Omega_\text{max}$ is almost indistinguishable from that of the standard protocol in the infinite blockade limit, confirming that for sufficiently large blockade strength the NH protocol is still able to preserve the benign scaling of the logical error rate. In contrast, for $B = 3,\Omega_\text{max}$, the NH protocol exhibits an increased logical error rate and reduced circuit-level distance, indicating that only $\lceil d/4\rceil$ physical errors are required to induce a logical error—analogous to the behavior observed for the time-optimal protocol.

In summary, the main text’s results for the NH protocol remain valid in the regime of strong but finite blockade interaction, which is relevant for most current neutral atom quantum processors. 
Conversely, in the weak blockade limit, both the NH and time-optimal protocols fail to maintain favorable logical error rate scaling, indicating the need for alternative pulse optimization strategies.

\vspace{0.5cm}

\section{S6. Error analysis for the SIM protocol}
\subsection{S6.1. Fit of the logical error rates}
\label{subsec:SI_sim_protocol_fit}
\begin{figure}
    \centering
    \includegraphics[width=\linewidth]{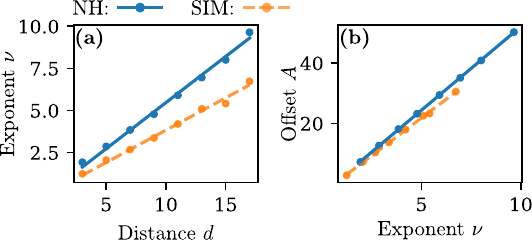}
    \caption{(a) Exponent $\nu$  with which $p_L$ increases with $\gamma$ as a function of code distance $d$ for the NH (blue, solid line) and the SIM (orange, dashed line) protocol. (b) Offset $A$ (see text) for the NH (blue, solid line) and the SIM (orange, dashed line) protocol.}
    \label{fig:SI_data_ancilla_fits}
\end{figure}

To understand the behavior of the crossing point between the NH and the SIM protocol [see Fig. 2(b) in the main text] we fit for each distance $d$ the logical error rate as $p_L = e^{A}\gamma^\nu$, with exponent $\nu$ and offset $A$. Fig.~\ref{fig:SI_data_ancilla_fits} shows a linear relationship $\nu = \alpha d$ between $\nu$ and $d$ and an affine relationship $A = \log(c) -\log(\gamma_{\mathrm{th}})\nu$ between $A$ and $\nu$, so that we obtain
\begin{equation}
    p_L = c\left(\frac{\gamma}{\gamma_{\mathrm{th}}}\right)^{\alpha d}.
\end{equation}
From fitting the curves in Fig.~\ref{fig:SI_data_ancilla_fits} we obtain 
\begin{align}
    \alpha = 0.55 \qquad \gamma_{\mathrm{th}} = 3.9\times 10^{-3} \qquad c=0.036
\end{align}
for the NH protocol and
\begin{align}
    \alpha' = 0.38 \qquad \gamma_{\mathrm{th}}' = 6.8\times 10^{-3} \qquad c'=0.044
\end{align}
for the SIM protocol.

The crossing point between protocols with different parameters $\gamma_{\mathrm{th}}, c, \alpha$ and $\gamma_{\mathrm{th}}', c', \alpha'$ is then given by 
\begin{equation}
   \gamma_{\times} = \left(c/c'\right)^{1/[(\alpha'-\alpha)d]}\left(\gamma_{\mathrm{th}}'^{\alpha'}/\gamma_{\mathrm{th}}^\alpha\right)^{1/(\alpha'-\alpha)}
\end{equation}
which converges to ${\gamma_{\mathrm{th}}'}^{\alpha'/(\alpha'-\alpha)}{\gamma_{\mathrm{th}}}^{\alpha/(\alpha-\alpha')}=1.1\times 10^{-3}/\Omega_{\mathrm{max}}$ as $d\rightarrow \infty$. This asymptotic crossing decay rate is shown by the black dashed line in Fig.~2(b) (main text).

\subsection{S6.2. Analytical Discussion}
\label{subsec:sim_gate_analytical_disucssion}
In the following we show that for the SIM protocol we have $\lambda_Q = \Theta(\gamma)$ for all 5-qubit Pauli errors $Q \neq I^{\otimes 5}$, i.e. that just one decay event can cause any Pauli error $Q$ on all 5 qubits on the plaquette. 

To show this it is sufficient to consider the decay of the ancilla qubit ($i=0$) at $t=2\pi/\Omega_{\mathrm{max}}$, i.e. exactly in the middle of the $2\pi$ pulse applied on the data qubits. Below we show that for any error $Q$ there exist $q$ and $k$ such that $\tr\left(QE_{0,k}^{(q)}\right) \neq 0$. Together with Eq.~\eqref{eq:SI_lambda_Q} this establishes $\lambda_Q = \Theta(\gamma)$.

Consider an initial state $\ket{a,d_1,d_2,d_3,d_4}$ of the ancilla and the data qubits, with $a,d_1,...,d_4 \in \{0,1\}$. Until the decay event, the evolution is given by 
\begin{equation}
    U(0,t)\ket{a,d_1,...,d_4} = \begin{cases}(-i)^{d_1+...+d_4}\ket{0,\bar{d_1},...\bar{d_4}} & \text{if } a=0 \\ -i\ket{r,d_1,...,d_4} & \text{if } a=1\end{cases}
\end{equation}
where for $x\in \{0,1\}$ we define $\bar{x}=0$ if $x=0$ and $\bar{x}=r$ if $x=1$. After the decay, i.e. after applying $L_0^{(q)}$ we are in the state
\begin{equation}
L_0^{(q)}U(0,t)\ket{a,d_1,...,d_4} = -i\delta_{a,1} \ket{q,d_1,...,d_4}
\end{equation}
where $\delta_{a,1} = 1$ if $a=1$ and $\delta_{a,1} = 0$ if $a=0$. Finally, we obtain $F_0^{(q)}$ as
\begin{equation}
    F_0^{(1)}\ket{a,0,...0} = -\delta_{a,1}\ket{r,0,...0}
\end{equation}
and
\begin{equation}
    F_0^{(q)}\ket{a,d_1...,d_4} = -i\delta_{a,1}(-i)^{d_1+...+d_4}\ket{q,\bar{d}_1,...,\bar{d}_4}
\end{equation}
if $q=0$ or $(d_1,...,d_4) \neq (0,...,0)$.

Now let $Q=Q_0\otimes...\otimes Q_4$ be a Pauli error. we choose $k_0=2$, for $j=1,...,4$ we choose $k_j=0$ if $Q_j\in \{X,Y\}$ and $k_j = 1$ if $Q_j \in \{I,Z\}$ , and $q = 0$ if $Q_0 \in \{X,Y\}$ and $q=1$ if $Q_0 \in \{I,Z\}$. Then we obtain
\begin{equation}
    D_kF_0^{(q)}\ket{a,d_1,...,d_4} = (-i)^5 \delta_{a,1}\delta_{d_1,1}...\delta_{d_4,1}\ket{q,k_1,...,k_4}
\end{equation}
and thus also
\begin{equation}
    E_{0,k}^{(q)}\ket{a,d_1,...,d_4} = - (-i)^5 \delta_{a,1}\delta_{d_1,1}...\delta_{d_4,1}\ket{q,k_1,...,k_4}.
\end{equation}
Hence we find
\begin{eqnarray}
    \tr\left(Q E_{0,k}^{(q)}\right) = - (-i)^5 \bra{1,...,1}Q\ket{q,k_1,...,k_4}
\end{eqnarray}
which is nonzero by the choice of $q$ and $k_1,...,k_4$.

\subsection{S6.3. Realistic experimental settings}
We conclude this section with a concise discussion of the SIM protocol’s error mechanisms, experimental challenges, and practical implementation strategies.

The SIM protocol implements a simultaneous $\pi$-$2\pi$-$\pi$ pulse on all five atoms in the plaquette, realizing a CZ$_4$ gate targeting the data and controlled by the ancilla (equivalent to four CZ gates applied in parallel). In general, multiqubit gates introduce correlated errors across several data qubits, which reduces the effective distance and undermines fault tolerance. This explains the findings of the main text, where $\lceil d/4\rceil$ errors--and not $\lceil d/2\rceil$--are found to be sufficient to trigger a logical error in this setting. The observed higher threshold instead reflects the shorter gate time compared to four sequential CZ gates along with the reduced stabilizer‑readout circuit depth (noting that the SIM protocol only requires one step instead of four to measure a single stabilizer).

In the main text, we have defined the SIM protocol assuming only data-ancilla blockade and using a discrete pulse sequence, namely, a $\pi$-$2\pi$-$\pi$. In the following, we discuss how this setting could be realized in current neutral atom experiments. First, the five‑qubit discrete pulse sequence could be made continuous and smooth by using two time-optimal global, continuous pulses (one for ancilla, one for all data) analogous to the optimization of the time‑optimal CZ gate pulse profile for two atoms (however, we note that a CZ$_4$ cannot be implemented with a single global pulse because it is not symmetric with respect to the five atoms \cite{three_qubit}). This is expected to yield a shorter gate duration and avoid equal‑amplitude constraints for all atoms which is challenging to realize in experiments. In addition, atomic level selection and geometry can be used to ensure that the data–ancilla interaction dominates over the data-data and ancilla-ancilla interactions, thus justifying the main text assumption of either zero or infinite blockade interaction. Finally, residual sensitivity to atomic motion at large but finite blockade may be mitigated by engineering robustness at pulse‑level at the cost of modestly longer gate durations.

\section{S7. All-to-All Blockade}
\begin{figure}
    \centering
    \includegraphics[width=\linewidth]{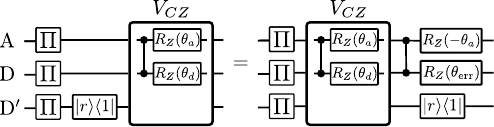}
    \caption{Propagation of a $\ket{r}\bra{1}$ error on a spectator data qubit. The error in the single qubit $Z$ rotation is $\theta_\mathrm{err} = - \theta_d$ under all-to-all blockade and $\theta_\mathrm{err} = 0$ under data-ancilla blockade.}
    \label{fig:SI_all_to_all_propagation}
\end{figure}

In the following we consider a second propagation mechanism for Rydberg leakage errors which is only present under all-to-all blockade, but not under data-ancilla blockade. For this, we consider how a $\ket{r}\bra{1}$ error on a \emph{spectator} data qubit $D'$, i.e. a qubit not involved in a $V_{CZ}$ gate between the ancilla qubit $A$ and a data qubit $D$, is propagated by this $V_{CZ}$ gate [Fig.~\ref{fig:SI_all_to_all_propagation}]. Under the all-to-all blockade model, the $\ket{r}\bra{1}$ error on $D'$ completely prevents the execution of the $V_{CZ}$ gate, leading to a $V_{CZ}^\dag = \mathrm{CZ}[R_Z(-\theta_a) \otimes R_Z(\theta_{\mathrm{err})}]${$V_{CZ}^\dagger = \mathrm{CZ}\cdot[R_Z(-\theta_a) \otimes R_Z(\theta_{\mathrm{err}})]$} error with $\theta_{\mathrm{err}} = -\theta_d$. We note that this is different from the data-ancilla blockade, where a Rydberg excitation of $D'$ still allows $D$ to be excited to the Rydberg state, so that $V_{CZ}$ acts like a single qubit $R_Z(\theta_d)$ gate on $D$, corresponding to $\theta_{\mathrm{err}} = 0$.

As we will show in the following, a $\ket{r}\bra{1}$ error on the first data qubit can lead to \emph{any} combination of Pauli error $Q = Q_0\otimes...\otimes Q_4$ with $Q_j \in \{I,Z\}$ on the 5 qubits of the plaquette through the propagation described in Fig.~\ref{fig:SI_all_to_all_propagation}, thus allowing $\lceil d/4 \rceil$ decay events to cause a logical error. The only exceptions are the cases $\theta_{\mathrm{err}}=0$ and $\theta_{\mathrm{err}}=\pi$, for which only errors with $Q_2=Q_3=Q_4$ are possible, so that up to $\lfloor d/2 \rfloor$ decay events can be corrected. Hence, the NH protocol can correct $\lfloor d/2\rfloor$ errors under the data-ancilla blockade ($\theta_{\mathrm{err}}=0$), but only $\lfloor d/4\rfloor$ errors under the all-to-all blockade ($\theta_{\mathrm{err}}=\pi/2$).

To determine the effect of a $\ket{r}\bra{1}$ error on the first data qubit, we calculate that by propagating through the three subsequent $V_{CZ}$ gates, the $\ket{1}\bra{r}$ error on the first data qubit causes an error
\begin{equation}
    E = (\ket{r}\bra{1})_1 \mathrm{CZ}^{(0,2)}\mathrm{CZ}^{(0,3)}\mathrm{CZ}^{(0,4)}E_{sq}
\end{equation}
where $(\ket{r}\bra{1})_1$ denotes a $\ket{r}\bra{1}$ error on qubit 1, $\mathrm{CZ}^{(i,j)}$ denotes a CZ gate between qubits $i$ and $j$ and
\begin{equation}
    E_{sq} = R_Z(-3\theta_d)\otimes I \otimes R_Z(\theta_{\mathrm{err}})^{\otimes 3}.
\end{equation}
The error $E$ can cause a Pauli error $Q$ if there is a Kraus operator $D$ of the channel $\mathcal{D}^{\otimes 5}$ such that $\tr(DEQ)\neq 0$. We now only consider $D = \Pi \otimes \ket{1}\bra{r} \otimes \Pi^{\otimes 3}$ and Pauli errors of the form $Q = Z^{z_0}\otimes ... \otimes Z^{z_4}$. We obtain for $a,d_1,...,d_4 \in \{0,1\}$ that
\begin{equation}
    DE\ket{a,d_1,...,d_4} = \delta_{d_1,1}e^{i\alpha}\ket{a,d_1,...,d_4}
\end{equation}
with 
\begin{equation}
    \alpha = \pi a(d_2+d_3+d_4) -3\theta_d a + \theta_{\mathrm{err}}(d_2+d_3+d_4).
\end{equation}
With that, we obtain
\begin{equation}
    \tr(DEQ) = (-1)^{z_1}\sum_{a=0}^1 e^{ia(\pi z_0 -3\theta_d)} \sum_{d_2,d_3,d_4=0}^1 e^{i\beta}
\end{equation}
with
\begin{equation}
    \beta = \sum_{j=2}^4 [\pi(z_j+a)+\theta_{\mathrm{err}}]d_j.
\end{equation}
We now obtain
\begin{equation}
    \sum_{d_2,d_3,d_4=0}^1 e^{i\beta} = \prod_{i=2}^4 \sum_{d=0}^1 e^{i[\pi(z_j+a)+\theta_{\mathrm{err}}]d}.
\end{equation}
The sum $\sum_{d=0}^1 \exp\{i[\pi(z_j+a)+\theta_{\mathrm{err}}]d\}$ vanishes if and only if $\pi(z_j+a)+\theta_{\mathrm{err}} = (2k+1)\pi$ for some integer $k$. Since $z_j, a \in \{0,1\}$ this is only possible if $\theta_{\mathrm{err}} \in \{0,\pi\}$. This establishes that for  $\theta_{\mathrm{err}} \notin \{0,\pi\}$ indeed every error $Q = Z^{z_0}\otimes Z^{z_4}$ is possible. In contrast, if $\theta_{\mathrm{err}} \in \{0,\pi\}$ then $\sum_{d=0}^1 \exp\{i[\pi(z_j+a)+\theta_{\mathrm{err}}]d\}$ only doesn't vanish for $z_j = a$ (for $\theta_{\mathrm{err}}=0$) or $z_j = a\oplus 1$ (for $\theta_{\mathrm{err}}=\pi$), where $\oplus$ denotes addition modulo 2. In either case, it follows that for $\tr(DEQ) \neq 0$ it is required that $z_2=z_3=z_4$.

\section{S8. The No-Phase (NP) protocol}
The No-Phase (NP) protocol [see Fig.~1(f)] is the shortest possible protocol that satisfies $c_1=1,c_2=c_3=c_4=0$ as well as $\theta_d\in\{0,\pi\}$, and thus restores the $\lceil d/2 \rceil$ scaling under data-data blockade. While we found this pulse using the quantum optimal control method of GRAPE \cite{khaneja_optimal_2005, jandura_time-optimal_2022}, it actually has a simple structure, which we describe in the following: On the ancilla qubit, two pulses with amplitude $\Omega_{\mathrm{max}}$ and duration $\tau_1$ are applied, where the second pulse has a phase shift of $\pi$ compared to the first pulse, and there is an idle time $T-2\tau_1$ in between them, where $T$ is the total gate duration. This two-pulse structure already ensures that $\ket{10} \mapsto \ket{10}$ under the NP gate. On the data qubit the laser pulse is instead always applied with amplitude $\Omega_{\mathrm{max}}$, but there are two phase jumps of $\pi$, switching the sign of $\Omega$, at times $\tau_2$ and $T-\tau_2$, with $\tau_2 = (T-2\pi/\Omega_{\mathrm{max}})/4$. Note that choice of $\tau_2$ ensures, that $\ket{01} \mapsto -\ket{01}$, so that $\theta_d=\pi/2$. By choosing $\tau_1 \approx 3.57/\Omega_{\mathrm{max}}$ and $T\approx9.20/\Omega_{\mathrm{max}}$ it is additionally achieved that $\ket{11} \mapsto \ket{11}$ and $c_1=1, c_2=c_3=c_4=0$.

\section{S9. Comparison between TO and $\pi$-$2\pi$-$\pi$ gate under all-to-all blockade}
\begin{figure}[t]
    \centering
    \includegraphics[width=\linewidth]{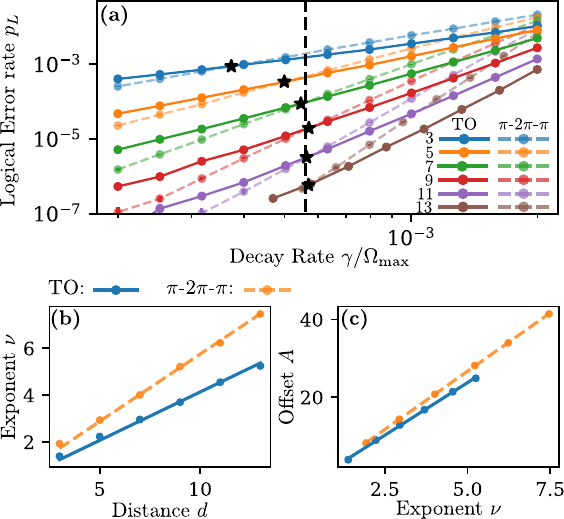}
    \caption{All-to-All blockade (a) Comparison of the logical error rate $p_L$ for the TO (solid curve) and the $\pi$-$2\pi$-$\pi$ (dashed curve) protocol. Stars mark the crossover points between the two protocols, the solid dashed line marks the value $\gamma_\times$ for which both protocols give the same $p_L$ in the $d\rightarrow\infty$ limit. (b) Exponent $\nu$  with which $p_L$ increases with $\gamma$ as a function of code distance $d$ for the TO (blue, solid line) and the $\pi$-$2\pi$-$\pi$ (orange, dashed line) protocol. (c) Offset $A$ (see text) for the TO (blue, solid line) and the $\pi$-$2\pi$-$\pi$ (orange, dashed line) protocol.}
    \label{figv2:SI_all_to_all_crossover}
\end{figure}

In the following we compare the TO and the $\pi$-$2\pi$-$\pi$ protocols, two of the most popular protocols in experimental implementations, under the all-to-all blockade. As discussed in the main text, the TO protocol is only able to correct $\lfloor d/4 \rfloor$ decay events, while the $\pi$-$2\pi$-$\pi$ can correct $\lfloor d/2 \rfloor$ decay events. However, the TO protocol has an almost 50\% smaller Rydberg time than the $\pi$-$2\pi$-$\pi$ protocol.

Fig.~\ref{figv2:SI_all_to_all_crossover}(a) shows the logical error rate untder the data-ancilla blockade for the TO and the $\pi$-$2\pi$-$\pi$ protocol for distances between $d=3$ and $d=13$ and $10^{-4} \leq \gamma/\Omega_{\mathrm{max}} \leq 2\times 10^{-3}$. For large decay rates $\gamma$ the TO protocol has the lower logical error rate, due to the smaller Rydberg time. However, for small decay rates $\gamma$ the $\pi$-$2\pi$-$\pi$ protocol has the lower logical error rate, due to its ability to correct $\lfloor d/2 \rfloor$ decay events. The black stars mark the points where both protocols have the same logical error rates. With increasing distance $d$ the crossing point moves to larger decay rates $\gamma$, but converges to a finite values as $d\rightarrow \infty$.

To determine the crossing point $\gamma_\times$ we proceed analogously to Sec~\ref{subsec:SI_sim_protocol_fit} and fit a linear relationship $\nu = \alpha d$ between $\nu$ and $d$ [\ref{figv2:SI_all_to_all_crossover}(b)] and  an affine relationship $A = \log(c) -\log(\gamma_{\mathrm{th}})\nu$ between $A$ and $\nu$ [\ref{figv2:SI_all_to_all_crossover}(a)]. From the fits in Fig~\ref{figv2:SI_all_to_all_crossover}(b/c) we extract
\begin{equation}
    \alpha = 0.41 \qquad \gamma_{\mathrm{th}} = 4.5\times 10^{-3} \qquad c=0.038
\end{equation}
for the TO protocol and
\begin{equation}
    \alpha' = 0.58 \qquad \gamma_{\mathrm{th}}' = 2.5\times 10^{-3} \qquad c'=0.038
\end{equation}
for the $\pi$-$2\pi$-$\pi$ protocol. This gives an asymptotic crossing point ${\gamma_{\mathrm{th}}'}^{\alpha'/(\alpha'-\alpha)}{\gamma_{\mathrm{th}}}^{\alpha/(\alpha-\alpha')}=5.6\times 10^{-3}/\Omega_{\mathrm{max}}$ as $d\rightarrow \infty$, comparable to current experimental error rates $10^{-3}\Omega_{\mathrm{max}} \lesssim \gamma \lesssim 10^{-4}\Omega_{\mathrm{max}}$.

\section{S10. Effect of Black-Body Radiation}

In the following, we discuss leakage errors by stimulated absorption or emission of black body radiation (BBR), which can cause excitations to a \emph{different} Rydberg state. While not captured in our model, we show below that in the presence of BBR decay the protocols developed in this work perform similarly to, and sometimes better than, the standard time-optimal protocol.

\begin{figure}[t]
    \centering
    \includegraphics[width=\linewidth]{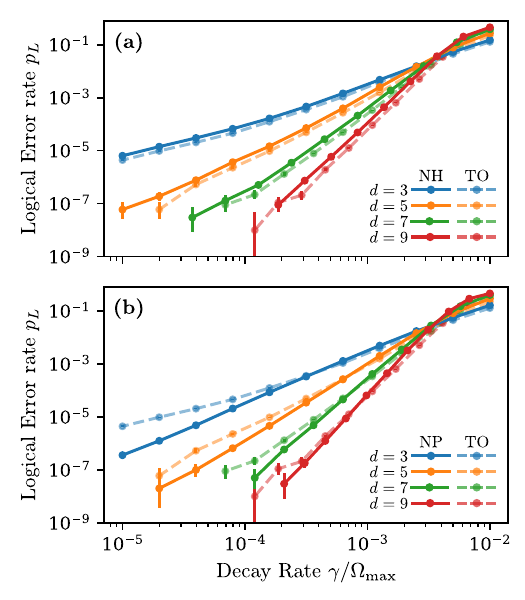}
    \caption{Data-ancilla blockade with BBR decays. (a) TO (dashed lines) and NH (solid lines) protocols perform comparably and both fail to maintain strict fault tolerance in presence of BBR decays. (b) NP protocol (solid lines), here defined in the data-ancilla blockade setting, instead preserve the favorable scaling of the logical error rate and hence outperforms both TO (dashed lines) and NH (not shown) protocols.}
    \label{figv2:SI_da_bbr}
\end{figure}

To account for BRR errors, we now model the atom as a four-level system, with computational states $\ket{0}$ and $\ket{1}$, auxiliary Rydberg state $\ket{r}$, and an additional Rydberg state $\ket{r'}$ which is only populated via BBR decay. In case of BBR decay, we assume that the population is transferred from $\ket{r}$ to $\ket{r'}$ and no decay to $\ket{0}$ or $\ket{1}$ can occur. Only when all the Rydberg population is decayed back to the computational states, also the additional Rydberg state $\ket{r'}$ is forced back to $\ket{0}$ or $\ket{1}$ with equal branching. Again, we recall that decays to low-lying non-computational states are operationally equivalent to decays to $\ket{0}$. We show in Fig.~\ref{figv2:SI_da_bbr}(a)--(b) surface code numerical simulations for TO, NH and NP protocols including BBR decays in the data-ancilla blockade setting. Fig.~\ref{figv2:SI_da_bbr}(a) shows that the NH protocol now behaves similarly to the TO protocol: It fails to maintain fault tolerance (i.e. $\lceil d/4\rceil$ physical errors suffice to introduce a logical error instead of $\lceil d/2\rceil$) and offers slightly larger logical error rates than the TO protocol due to the moderately larger pulse duration. We find instead that the NP protocol--in the main text defined in the all-to-all blockade regime, but here extended to the data-ancilla setting--preserves fault tolerance and offer the lowest logical error rates compared to the other protocols. These effects follow from the protocols' pulse designs. The NH protocol suppresses hopping of an excitation $\ket{r}\bra{1}$ between data qubits, but this protection fails if BBR populates a different Rydberg level $\ket{r'}$. A $\ket{r'}\bra{1}$ excitation on a spectator qubit can induce a data–data blockade that prevents the execution of $V_\text{CZ}$. Experimentally, BBR can populate higher‑lying or nearby $P$ states with longer‑range or anisotropic interactions that can perturb neighboring atoms. By contrast, the NP protocol is immune by design to these types of errors and restores the favorable logical‑error scaling in the presence of BBR processes. We do not show results for the all-to-all blockade setting, but we have numerically verified that the NP protocol still succeeds to maintain the benign scaling of the logical error rate and thus offers the lowest logical error rate compared to standard protocols.

Thus, we have shown that in presence of BBR decays that can populate neighboring Rydberg states, the novel NP protocol designed in this work performs best against all other pulse protocols, both in data-ancilla and all-to-all blockade settings. Missing atoms or population in low‑lying non‑computational states prevent subsequent CZ operations and are therefore operationally equivalent to qubits in $\ket{0}$. The same effect can be reproduced in stationary atom arrays by either waiting after each gate or by switching on and off the traps to convert these errors into atom loss. Hence, BBR‑induced decays within a stabilizer round can be absorbed into our noise model by increasing the effective decay probability to $\ket{0}$ and do not undermine the conclusion that the advantages of the proposed gate protocols are readily accessible in current neutral‑atom platforms.

\vspace{0.5cm}

\section{S11. Atom loss and inter-round leakage propagation}
We have already discussed both in main text and previous supplemental sections that qubit loss within a single stabilizer round (from atom loss or decay to long‑lived non‑computational states) can be incorporated in our model as an increased effective leakage to $|0\rangle$, since $|0\rangle$ is dark to the Rydberg drive and thus prevents CZ execution in the same way as a missing atom. Without dedicated detection and correction, these errors can persist and propagate across multiple QEC rounds, severely degrading the QEC performance. Inter‑round leakage propagation is not directly captured within our noise model. Below we summarize circuit‑ and hardware‑level mitigation strategies to prevent such inter‑round propagation and discuss how these operations can be incorporated into our framework to yield a comprehensive model of leakage in neutral atom quantum processors.

\emph{Leakage reduction units.} It was first shown in \cite{aliferis2007} and afterwards in \cite{fowler2013,suchara2015,gosh2015} that that leakage‑resilient QEC and recovery of a threshold can be achieved by inserting leakage reduction units (LRUs) after each stabilizer round. LRUs are small fault‑tolerant circuits that entangle data qubits with new ancillary qubits to detect and, when needed, reload leaked qubits. These gadgets introduce extra complexity and overheads, but make leakages correctable by converting them into regular Pauli errors. There exist two main types of LRUs: the standard LRU and the teleportation-based LRU. The former entangles the presence or absence of a data qubits with a new ancilla qubit and reloads a fresh data if necessary. The latter consists in teleporting the states of old and ``hot" data qubits into a new layer of fresh qubits. The performance of the rotated surface code supplemented with LRUs has been first studied at circuit-level in \cite{Perrin_2025,baranes2025} under a mixture of depolarizing and loss errors. The teleportation-based LRU is found to outperform the standard LRU, and logical error rate and QEC threshold are shown to improve as the loss fraction increases due to the possibility of converting losses into ``delayed" erasures--namely errors with known locations--at the end of each round via loss‑aware decoding. LRUs have been first experimentally demonstrated in a neutral atom experiment in \cite{chow_circuit-based_2024}.
More recent experiments have shown that neutral‑atom platforms can detect and correct (via reloading) mid‑circuit atom loss while preserving neighboring‑qubit coherence \cite{atom_computing2025}. Teleportation‑based LRUs have been implemented to mitigate losses while maintaining constant entropy throughout computation, thereby enabling repeated, sub‑threshold QEC in reconfigurable neutral‑atom quantum processors \cite{bluvstein2025architecturalmechanismsuniversalfaulttolerant}.

\emph{Erasure conversion, loss detection, and optical re-pumping.} Numerous operations to herald and correct leakages at the hardware-level have also been demonstrated in recent theoretical and experimental works. In alkali earth(-like) neutral atom platforms, leakages can be forced to return the atom to the true ground state rather than to the metastable qubit states and converted into erasures via fluorescence-imaging of the ground state after each gate. This operation leaves the qubit state untouched, while allowing to herald possible leakages events. Atoms involved in the erased CZ gate can be replaced with fresh ones, thus practically converting large fractions of leakages into erasures \cite{wu_erasure_2022,ma_high-fidelity_2023}. This erasure conversion protocol has been show to significantly improve the error correction performance of many QEC codes, including surface and quantum LDPC codes \cite{sahay_high_2023,ldpc_erasures2025}. Recent experiments with alkali earth(-like) neutral atom qubits have shown how loss-detectable spin measurement can be used to detect leakages from the Rydberg states and hence increase the fraction of detected erasure errors after qubit detection is performed \cite{senoo2025highfidelityentanglementcoherentmultiqubit}. Such delayed heralding of erasure errors can be used to detect leakages and prevent their propagation across multiple stabilizer measurement rounds in QEC experiments. Even more recently, a new non-destructive spin-resolved readout scheme has been experimentally demonstrated with alkali atoms \cite{bluvstein2025architecturalmechanismsuniversalfaulttolerant}. Such readout scheme uses a one-dimensional state-selective optical lattice to enable both loss detection, atom retention and hence repeated QEC cycles via atom reloading. Hyperfine leakages--namely leakages to the wrong hyperfine ground states possibly affecting alkali atoms--can instead be mitigated via optical re-pumping \cite{cong_hardware-efficient_2022}. However, recent experiment have found that hyperfine leakages by gate errors are naturally largely suppressed and loss-converted due to anti-trapping of highly-excited states \cite{bluvstein2025architecturalmechanismsuniversalfaulttolerant}.

In conclusion, these observations indicate that qubit‑loss errors require dedicated mitigation strategies to prevent inter‑round propagation. This involves extra operations and overheads both at hardware- and circuit-level. Once inter‑round loss is addressed, leakage confined to individual stabilizer rounds can be treated with the methodology presented in this work. Accordingly, when complemented by suitable loss‑detection and correction protocols, our noise model captures the dominant leakage mechanisms affecting neutral‑atom processors.

\end{document}